\documentclass[10pt, conference, compsocconf]{IEEEtran}
\ifCLASSINFOpdf
\else
\fi
\usepackage[caption=true,font=footnotesize]{caption}
\usepackage{mathptmx}
\usepackage[dvips]{graphicx}
\graphicspath{{figures/}}
\DeclareGraphicsExtensions{.eps}
\usepackage{color}
\definecolor{grey}{rgb}{0.95, 0.95, 0.95}
\usepackage[final]{listings}
\usepackage{array}

\begin{document}
\title{LIKWID: A lightweight performance-oriented tool suite for x86 multicore environments}

\author{\IEEEauthorblockN{Jan Treibig, Georg Hager, Gerhard Wellein}
\IEEEauthorblockA{Erlangen Regional Computing Center (RRZE)\\
University of Erlangen-Nuremberg\\
Erlangen, Germany\\
Email: jan.treibig@rrze.uni-erlangen.de}}
\maketitle
\begin{abstract}
    Exploiting the performance of today's processors requires
    intimate knowledge of the microarchitecture as well as an awareness
    of the ever-growing complexity in thread and cache topology.
    LIKWID is a set of command-line utilities that addresses four key problems: 
    Probing the thread and cache topology
    of a shared-memory node, enforcing thread-core affinity on a program, 
    measuring performance counter metrics, and toggling hardware
    prefetchers. An API for using the performance counting features from user code
    is also included. We clearly state the differences to the widely used PAPI 
    interface. To demonstrate the capabilities of the tool set we show the
    influence of thread pinning on performance using the well-known OpenMP
    STREAM triad benchmark, and use the affinity and hardware counter tools
    to study the performance of a stencil code specifically
    optimized to utilize shared caches on multicore chips.
\end{abstract}

\section{Introduction}

\begin{figure}[tb]\centering
    \includegraphics*[width=0.9\linewidth]{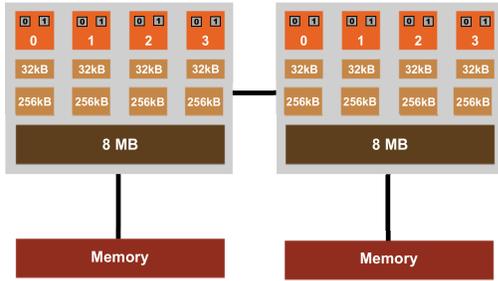}
    \caption{Thread and cache topology of an Intel Nehalem EP multicore dual-socket node}
    \label{fig:nehalem_socket}
\end{figure}

Today's multicore x86 processors bear multiple complexities when aiming for
high performance. Conventional performance tuning tools like Intel VTune,
OProfile, CodeAnalyst, OpenSpeedshop, etc., require a lot of experience in order
to get sensible results. For this reason they are usually unsuitable for the
scientific user, who would often be satisfied with a rough overview of the
performance properties of their application code. Moreover, advanced tools
often require kernel patches and additional software components, which makes
them unwieldy and bug-prone. Additional confusion arises with the complex
multicore, multicache, multisocket structure of modern systems 
(see Fig.~\ref{fig:nehalem_socket}); users
are all too often at a loss about how hardware thread 
IDs are assigned to resources like cores, caches, sockets and NUMA
domains. Moreover, the technical details of how threads and processes
are bound to those resources vary strongly across compilers and MPI 
libraries.

LIKWID (``Like I Knew What I'm Doing'') is a set of easy to use command line
tools to support optimization. It is targeted towards performance-oriented
programming in a Linux environment, does not require any kernel patching, and
is suitable for Intel and AMD processor architectures. Multithreaded and even
hybrid shared/distributed-memory parallel code is supported. It comprises 
the following tools:
\begin{itemize}
\item \verb.likwid-features. can display and alter the state of the
  on-chip hardware prefetching units in Intel x86 processors. 
\item \verb.likwid-topology. probes the hardware thread and cache
  topology in multicore, multisocket nodes. Knowledge like this is
  required to optimize resource usage like, e.g., shared caches and
  data paths, physical cores, and ccNUMA locality domains, in parallel
  code.
\item \verb.likwid-perfCtr. measures performance counter metrics over
  the complete runtime of an application or, with support from a
  simple API, between arbitrary points in the code. Counter
  multiplexing allows the concurrent measurement of a large number of
  metrics, larger than the (usually small) number of available
  counters. Although it is possible to specify the full,
  hardware-dependent event names, some predefined event sets simplify
  matters when standard information like memory bandwidth or Flop counts
  is needed.
\item \verb.likwid-pin. enforces thread-core affinity
  in a multi-threaded application ``from the outside,'' i.e., without
  changing the source code. It works with all threading models that
  are based on POSIX threads, and is also compatible with hybrid 
  ``MPI+threads'' programming. Sensible use of likwid-pin requires
  correct information about thread numbering and cache topology, 
  which can be delivered by likwid-topology (see above).
\end{itemize}
Although the four tools may appear to be partly unrelated, they solve the
typical problems application programmers have when porting and running
their code on complex multicore/multisocket environments.
Hence, we consider it a natural idea to provide them as a single
tool set.

This paper is organized as follows. Section~\ref{sec:tools} describes 
the four tools in some detail and gives hints for typical use.
In Section~\ref{sec:papi} we briefly compare LIKWID to the PAPI feature set.
Section~\ref{sec:cases} demonstrates the use of LIKWID in three
different case studies, and Section~\ref{sec:conc} gives a summary and
an outlook to future work.

\section{Tools}\label{sec:tools}

LIKWID only supports x86-based processors.  Given the strong prevalence
of those architectures in the HPC market (e.g., 90\% of all systems in
the latest Top 500 list are of x86 type) we do not consider this a severe
limitation. In other areas like, e.g., workstations or desktops,
the x86 dominance is even larger. 

In the following we describe the four tools in detail.

\subsection{likwid-perfCtr}

\begin{figure}[tb]\centering
    \includegraphics*[width=0.9\linewidth]{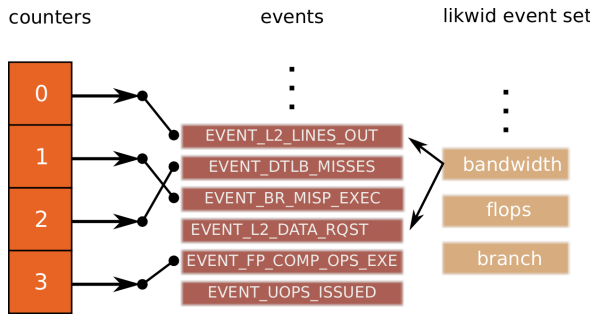}
    \caption{likwid-perfCtr: Interaction between event sets, hardware events and performance counters.}
    \label{fig:perfCtr}
\end{figure}
Hardware-specific optimization requires an intimate knowledge of the
microarchitecture of a processor and the characteristics of the code. While
many problems can be solved with profiling, common sense, and runtime measurements,
additional information is often useful to get a complete picture. 

Performance counters are facilities to count hardware events during
code execution on a processor. Since this mechanism is implemented directly in
hardware there is no overhead involved. All modern processors provide hardware
performance counters, but their primary purpose is to support computer
architects during the implementation phase. Still they are also attractive for
application programmers, because they allow an in-depth view on what happens on
the processor while running applications. There are generally two options for using 
hardware performance counter data: Either event counts are collected over
the runtime of an application process (or probably restricted to certain
code parts via an appropriate API), or overflowing hardware counters can generate
interrupts, which can be used for IP or call-stack sampling. The latter option
enables a very fine-grained view on a code's resource requirements (limited
only by the inherent statistical errors). However, the first option is
sufficient in many cases and also practically overhead-free. This is why
it was chosen as the underlying principle for likwid-perfCtr.

The probably best known and widespread existing tool is the PAPI library \cite{PAPI},
for which we provide a detailed comparison to likwid-perfCtr in Section~\ref{sec:comp_papi}. A
lot of research is targeted towards using performance counter data for
automatic performance analysis and detecting potential performance bottlenecks
\cite{1213183,1418278,DBLP:conf/parco/GerndtFK05}. However, those solutions are
often too unwieldy for the common user, who would prefer a quick overview as a first step
in performance analysis. A key design goal for likwid-perfCtr was ease of
installation and use, minimal system requirements (no additional kernel modules
and patches), and --- at least for basic functionality --- no changes to the
user code.
A prototype for the development of likwid-perfCtr is the SGI tool
``perfex,'' which was available on  MIPS-based IRIX machines as part
of the ``SpeedShop'' performance suite. Cray provides a
similar, PAPI-based tool (craypat) on their systems.  likwid-perfCtr
offers comparable or improved functionality with regard to hardware
performance counters on x86 processors, and is available as open
source.

Hardware performance counters are controlled and accessed using
processor-specific hardware registers (also called \emph{model specific registers}
(MSR)). likwid-perfCtr uses the Linux ``msr'' module to modify the MSRs from user
space. The msr module is available in all Linux distributions with a 2.6 Linux kernel
and implements the read/write access to MSRs based on device files.

likwid-perfCtr is a command line tool that can be used as a wrapper to an
application. It allows simultaneous measurements on multiple cores.  Events
that are shared among the cores of a socket (this pertains to the ``uncore''
events on Core i7-type processors) are supported via ``socket locks,'' which
enforce that all uncore event counts are assigned to one thread per socket.
Events are specified on the command line, and the number of events to count
concurrently is limited by the number of performance counters on the CPU. These
features are available without any changes in the user's source code.  A small
instrumentation (``marker'') API allows one to restrict measurements to certain
parts of the code (named regions) with automatic accumulation over all regions
of the same name. An important difference to most existing performance tools is
that event counts are strictly core-based instead of process-based: Everything
that runs and generates events on a core is taken into account; no attempt is
made to filter events according to the process that caused them. The user is
responsible for enforcing appropriate affinity to get sensible results.
 This could be achieved via likwid-pin (see below for more information):
\begin{lstlisting}[basicstyle=\footnotesize\ttfamily]
$ likwid-perfCtr -c 1 \
   -g SIMD_COMP_INST_RETIRED_PACKED_DOUBLE:PMC0,\
      SIMD_COMP_INST_RETIRED_SCALAR_DOUBLE:PMC1 \
         likwid-pin -c 1 ./a.out
\end{lstlisting}
(See below for typical output in a more elaborate setting.)
In this example, the computational double precision 
packed and scalar SSE  retired instruction counts on an Intel Core 2 
processor are assigned to performance
counters 0 and 1 and measured on core 1 over the
duration of \verb!a.out!'s runtime. The \verb.likwid-pin. command
is used here to bind the process to this core.
As a side effect, it becomes possible to use likwid-perfCtr as
a monitoring tool for a complete shared-memory node, just by specifying
all cores for measurement and, e.g., ``\verb.sleep.'' as 
an application:
\begin{lstlisting}[basicstyle=\footnotesize\ttfamily]
$ likwid-perfCtr -c 0-7 \
   -g SIMD_COMP_INST_RETIRED_PACKED_DOUBLE:PMC0,\
      SIMD_COMP_INST_RETIRED_SCALAR_DOUBLE:PMC1 \
         sleep 1
\end{lstlisting}

Apart from naming events as they are documented in the vendor's
manuals, it is also possible to use preconfigured \emph{event sets} (groups)
with derived metrics. This provides a simple abstraction layer in
cases where standard information like memory bandwidth, Flops per
second, etc., is sufficient:
\begin{lstlisting}[basicstyle=\footnotesize\ttfamily]
$ likwid-perfCtr -c 0-3  \
   -g FLOPS_DP  ./a.out
\end{lstlisting}
At the time of writing, the following event sets are defined:
\begin{center}
\begin{tabular}{r|m{5.5cm}}
\bfseries Event set & \bfseries Function \\\hline
\ttfamily FLOPS\_DP & Double Precision MFlops/s\\
\ttfamily FLOPS\_SP & Single Precision MFlops/s\\
\ttfamily L2 & L2 cache bandwidth in MBytes/s\\
\ttfamily L3 & L3 cache bandwidth in MBytes/s\\
\ttfamily MEM & Main memory bandwidth in MBytes/s\\
\ttfamily CACHE & L1 Data cache miss rate/ratio\\
\ttfamily L2CACHE & L2 Data cache miss rate/ratio\\
\ttfamily L3CACHE & L3 Data cache miss rate/ratio\\
\ttfamily DATA & Load to store ratio\\
\ttfamily BRANCH & Branch prediction miss rate/ratio\\
\ttfamily TLB & Translation lookaside buffer miss rate/ratio
\end{tabular}
\end{center}
The event groups are partly inspired from a technical report published by AMD \cite{AMD}.
We try to provide the same preconfigured event groups on all supported 
architectures, as long as the native events support them. This allows the 
beginner
to concentrate on the useful information right away, without the 
need to look up events in the manuals (similar to PAPI's
high-level events). 

The interactions between event sets, hardware events, and performance
counters are illustrated in Fig.~\ref{fig:perfCtr}. In the usage
scenarios described so far there is no interference of likwid-perfCtr
while user code is being executed, i.e., the overhead is very small
(apart from the unavoidable API call overhead in marker mode).  If the
number of events is larger than the number of available counters, this
mode of operation requires running the application more than once. For
ease of use in such situations, likwid-perfCtr also supports a
\emph{multiplexing mode}, where counters are assigned to several event
sets in a ``round robin'' manner.  On the downside, short-running
measurements will then carry large statistical errors. Multiplexing is
supported in wrapper and marker mode.

The following example illustrates the use of the marker API in a serial
program with two named regions (``\verb.Main.'' and ``\verb.Accum.''):
\begin{lstlisting}[basicstyle=\footnotesize\ttfamily]
#include <likwid.h>
...
int coreID = likwid_processGetProcessorId();
printf("Using likwid\n");
likwid_markerInit(numberOfThreads,numberOfRegions);
int MainId  = likwid_markerRegisterRegion("Main");
int AccumId = likwid_markerRegisterRegion("Accum");

likwid_markerStartRegion(0, coreID);
// measured code region
likwid_markerStopRegion(0, coreID, MainId);

for (j = 0; j < N; j++)
{
   likwid_markerStartRegion(0, coreID);
   // measured code region
   likwid_markerStopRegion(0, coreID, AccumId);
}

likwid_markerClose();
\end{lstlisting}
Event counts are automatically accumulated on multiple calls.  Nesting or
partial overlap of code regions is not allowed. The API requires specification of a
thread ID (0 for one process only in the example) and the core ID of the
thread/process. The likwid API provides simple functions to determine the core
ID of processes or threads.  The following listing shows the output of
likwid-perfCtr after measurement of the \verb.FLOPS_DP. event group on four
cores of an Intel Core 2 Quad processor in marker mode with two named regions
(``\verb.Init.'' and ``\verb.Benchmark.,'' respectively):
\begin{lstlisting}
$ likwid-perfCtr -c 0-3 -g FLOPS_DP -m ./a.out
-------------------------------------------------------------
CPU type:       Intel Core 2 45nm processor
CPU clock:      2.83 GHz
-------------------------------------------------------------
Measuring group FLOPS_DP
-------------------------------------------------------------
%Region: Init% 
+--------------------------------------+--------+--------+--------+--------+
|                Event                 | core 0 | core 1 | core 2 | core 3 |
+--------------------------------------+--------+--------+--------+--------+
|          INSTR_RETIRED_ANY           | 313742 | 376154 | 355430 | 341988 |
|        CPU_CLK_UNHALTED_CORE         | 217578 | 504187 | 477785 | 459276 |
| SIMD_COMP_INST_RETIRED_PACKED_DOUBLE |   0    |   0    |   0    |   0    |
| SIMD_COMP_INST_RETIRED_SCALAR_DOUBLE |   1    |   1    |   1    |   1    |
+--------------------------------------+--------+--------+--------+--------+
+-------------+-------------+-------------+-------------+-------------+
|   Metric    |   core 0    |   core 1    |   core 2    |   core 3    |
+-------------+-------------+-------------+-------------+-------------+
| Runtime [s] | 7.67906e-05 | 0.000177945 | 0.000168626 | 0.000162094 |
|     CPI     |  0.693493   |   1.34037   |   1.34424   |   1.34296   |
| DP MFlops/s |  0.0130224  | 0.00561973  | 0.00593027  | 0.00616926  |
+-------------+-------------+-------------+-------------+-------------+
%Region: Benchmark% 
+-----------------------+-------------+-------------+-------------+-------------+
|           Event       |   core 0    |   core 1    |   core 2    |   core 3    |
+-----------------------+-------------+-------------+-------------+-------------+
| INSTR_RETIRED_ANY     | 1.88024e+07 | 1.85461e+07 | 1.84947e+07 | 1.84766e+07 |
| CPU_CLK_UNHALTED_CORE | 2.85838e+07 | 2.82369e+07 | 2.82429e+07 | 2.82066e+07 |
| SIMD_...PACKED_DOUBLE |  8.192e+06  |  8.192e+06  |  8.192e+06  |  8.192e+06  |
| SIMD_...SCALAR_DOUBLE |      1      |      1      |      1      |      1      |
+-----------------------+-------------+-------------+-------------+-------------+
+-------------+-----------+------------+------------+------------+
|   Metric    |  core 0   |   core 1   |   core 2   |   core 3   |
+-------------+-----------+------------+------------+------------+
| Runtime [s] | 0.0100882 | 0.00996574 | 0.00996787 | 0.00995505 |
|     CPI     |  1.52023  |  1.52252   |  1.52708   |  1.52661   |
| DP MFlops/s |  1624.08  |  1644.03   |  1643.68   |   1645.8   |
+-------------+-----------+------------+------------+------------+
\end{lstlisting}
Note that the \verb.INSTR_RETIRED_ANY. and \verb.CPU_CLK_UNHALTED_CORE. events
are always counted (using two unassignable ``fixed counters'' on the Core 2
architecture), so that the derived \verb.CPI. metric (``cycles per instruction'')
is easily obtained.

The following architectures are supported at the time of writing:
\begin{itemize}
    \item Intel Pentium M (Banias, Dothan)
    \item Intel Atom
    \item Intel Core 2 (all variants)
    \item Intel Nehalem (all variants, including uncore events) 
    \item Intel Westmere
    \item AMD K8 (all variants) 
    \item AMD K10 (Barcelona, Shanghai, Istanbul) 
\end{itemize}

\subsection{likwid-topology}

Multicore/multisocket machines exhibit complex topologies, and this trend will
continue with future architectures. Performance programming requires in-depth
knowledge of cache and node topologies, e.g., about which caches are shared between
which cores and which cores reside on which sockets. The Linux kernel numbers
the usable cores and makes this information accessible in \verb./proc/cpuinfo.. Still
how this numbering maps to the node topology depends on BIOS settings and may
even differ for otherwise identical processors. The processor and cache
topology can be queried with the \verb+cpuid+ machine instruction. likwid-pin
is based directly on the data provided by \verb+cpuid+. It extracts machine
topology in an accessible way and can also report on cache characteristics. The
thread topology is determined from the APIC (Advanced Programmable Interrupt
Controller) ID. Starting with the Nehalem processor, Intel introduced a new
\verb+cpuid+ leaf (0xB) to account for today's more complex multicore chip
topologies. Older Intel and AMD processors both have different methods to
extract this information, all of which are supported by likwid-topology.
Similar considerations apply for determining the cache topology. Starting with
the Core 2 architecture Intel introduced the \verb+cpuid+ leaf 0x4
(deterministic cache parameters), which allows to extract the cache
characteristics and topology in a systematic way.  On older Intel processors
the cache parameters where provided by means of a lookup table (\verb+cpuid+
leaf 0x2). AMD again has its own \verb+cpuid+ leaf for the cache parameters.
The core functionality of likwid-topology is implemented in a C module,
which can also be used as a library to access the information from
within an application.

likwid-topology outputs the following information:
\begin{itemize}
    \item Clock speed
    \item Thread topology (which hardware threads map to which physical resource)
    \item Cache topology (which hardware threads share a cache level)
    \item Extended cache parameters for data caches.
\end{itemize}
The following output was obtained on an Intel Nehalem EP Westmere
processor and includes extended cache information:
\begin{lstlisting}
$ likwid-topology -c
-------------------------------------------------------------
CPU name:       Unknown Intel Processor 
CPU clock:      2.93 GHz 

*************************************************************
Hardware Thread Topology
*************************************************************
Sockets:                2 
Cores per socket:       6 
Threads per core:       2 
-------------------------------------------------------------
HWThread        Thread          Core            Socket
0               0               0               0
1               0               1               0
2               0               2               0
3               0               8               0
4               0               9               0
5               0               10              0
6               0               0               1
7               0               1               1
8               0               2               1
9               0               8               1
10              0               9               1
11              0               10              1
12              1               0               0
13              1               1               0
14              1               2               0
15              1               8               0
16              1               9               0
17              1               10              0
18              1               0               1
19              1               1               1
20              1               2               1
21              1               8               1
22              1               9               1
23              1               10              1
-------------------------------------------------------------
Socket 0: ( 0 12 1 13 2 14 3 15 4 16 5 17 )
Socket 1: ( 6 18 7 19 8 20 9 21 10 22 11 23 )
-------------------------------------------------------------

*************************************************************
Cache Topology
*************************************************************
Level:   1
Size:    32 kB
Type:    Data cache
Associativity:   8
Number of sets:  64
Cache line size: 64
Inclusive cache
Shared among 2 threads
Cache groups:   ( 0 12 ) ( 1 13 ) ( 2 14 ) ( 3 15 ) ( 4 16 )
( 5 17 ) ( 6 18 ) ( 7 19 ) ( 8 20 ) ( 9 21 ) ( 10 22 ) ( 11 23 )
-------------------------------------------------------------
Level:   2
Size:    256 kB
Type:    Unified cache
Associativity:   8
Number of sets:  512
Cache line size: 64
Inclusive cache
Shared among 2 threads
Cache groups:   ( 0 12 ) ( 1 13 ) ( 2 14 ) ( 3 15 ) ( 4 16 )
( 5 17 ) ( 6 18 ) ( 7 19 ) ( 8 20 ) ( 9 21 ) ( 10 22 ) ( 11 23 )
-------------------------------------------------------------
Level:   3
Size:    12 MB
Type:    Unified cache
Associativity:   16
Number of sets:  12288
Cache line size: 64
Non Inclusive cache
Shared among 12 threads
Cache groups:   ( 0 12 1 13 2 14 3 15 4 16 5 17 )
( 6 18 7 19 8 20 9 21 10 22 11 23 )
-------------------------------------------------------------
\end{lstlisting}
One can also get an accessible overview of the node's cache and socket
topology in ASCII art (via the \verb.-g. option). 
The following listing fragment shows the output for the same
chip as above. Note that only one socket is shown
(belonging to the first L3 cache group above):
\begin{lstlisting}
+-------------------------------------------------------------+
| +-------+ +-------+ +-------+ +-------+ +-------+ +-------+ |
| | 0  12 | | 1  13 | | 2  14 | | 3  15 | | 4  16 | | 5  17 | |
| +-------+ +-------+ +-------+ +-------+ +-------+ +-------+ |
| +-------+ +-------+ +-------+ +-------+ +-------+ +-------+ |
| |  32kB | |  32kB | |  32kB | |  32kB | |  32kB | |  32kB | |
| +-------+ +-------+ +-------+ +-------+ +-------+ +-------+ |
| +-------+ +-------+ +-------+ +-------+ +-------+ +-------+ |
| | 256kB | | 256kB | | 256kB | | 256kB | | 256kB | | 256kB | |
| +-------+ +-------+ +-------+ +-------+ +-------+ +-------+ |
| +---------------------------------------------------------+ |
| |                           12MB                          | |
| +---------------------------------------------------------+ |
+-------------------------------------------------------------+
\end{lstlisting}

\subsection{likwid-pin}

\begin{figure}[tb]\centering
    \includegraphics*[width=\linewidth]{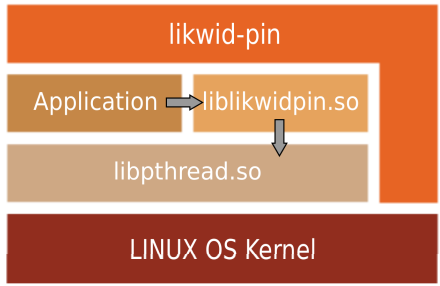}
    \caption{Basic structure of likwid-pin.}
    \label{fig:likwid-pin}
\end{figure}

Thread/process affinity is vital for performance. If topology information is
available, it is possible to pin threads according to the application's resource
requirements like bandwidth, cache sizes, etc. Correct pinning is even more
important on processors supporting SMT, where multiple hardware threads share resources
on a single core. likwid-pin supports thread affinity for all threading models that
are based on POSIX threads, which includes most OpenMP implementations. By
overloading the \verb+pthread_create+ API call with a shared library wrapper, 
each thread can be pinned in turn upon creation, working through a list of 
core IDs. This list, and possibly other parameters, are encoded in environment
variables that are evaluated when the library wrapper is first called. 
likwid-pin simply starts the user application with the library
preloaded.

The overall mechanism is illustrated in Fig.~\ref{fig:likwid-pin}.
No code changes are required, but the application must be dynamically linked.
This mechanism is independent of processor architecture, but the
way the compiled code creates application threads must be taken
into account: For
instance, the Intel OpenMP implementation  always runs \verb-OMP_NUM_THREADS+1-
threads but uses the first newly created thread as a management thread,
which should not be pinned. This knowledge must be conveyed to the 
wrapper library.  The following
example shows how to use likwid-pin with an OpenMP application compiled with
the Intel compiler:
\begin{lstlisting}[basicstyle=\small\ttfamily]
$ export OMP_NUM_THREADS=4
$ likwid-pin -c 0-3 %-t intel% ./a.out
\end{lstlisting}
Currently, POSIX threads, Intel OpenMP, and GNU (gcc) OpenMP are supported, 
and the latter is assumed as the default if no \verb.-t. switch is
given.
Other threading implementations are supported via a ``skip mask.'' This mask
is interpreted as a binary pattern and specifies which threads should not be 
pinned by the wrapper library (the explicit mask for Intel binaries would
by \verb.0x1.).  The skip mask
makes it possible to pin hybrid applications as well by skipping MPI shepherd
threads. For Intel-compiled binaries using the Intel MPI library,
the appropriate skip mask is \verb.0x3.:
\begin{lstlisting}[basicstyle=\small\ttfamily]
$ export OMP_NUM_THREADS=8
$ mpiexec -n 64 -pernode \
     likwid-pin -c 0-7 %-s 0x3% ./a.out
\end{lstlisting}
This would start 64 MPI processes on 64 nodes (via the \verb.-pernode.
option) with eight threads each, and not bind the first two newly
created threads.

In general, likwid-pin can be used as a
replacement for the \verb.taskset. tool, which cannot pin threads
individually. Note, however, that likwid-pin, in contrast to \verb.taskset.,
does not establish a Linux cpuset in which to run the application.

Some compilers have their own means for enforcing thread affinity. In order to
avoid interference effects, those mechanisms should be disabled when using
likwid-pin. In case of recent Intel compilers, this can be achieved by setting the
environment variable \verb+KMP_AFFINITY+ to \verb.disabled.
The current version of LIKWID does this automatically.

The big advantage of likwid-pin is its portable 
approach to the pinning problem, since the same tool can be used for all
applications, compilers, MPI implementations, and processor types. In 
Section~\ref{sec:case_1}
the usage model is analyzed in more detail on the example of the STREAM triad.

\subsection{likwid-features}

An important hardware optimization on modern processors is to hide data access
latencies by  hardware prefetching. Intel processors not only have a prefetcher
for main memory; several prefetchers are responsible for moving data
between cache levels. Often it is beneficial to know the influence of the
hardware prefetchers. In some situations turning off hardware prefetching 
even increases performance. On the Intel Core 2 processor this can be achieved
by setting bits in the \verb+IA32_MISC_ENABLE MSR+ register. likwid-features
allows viewing and altering the state of these bits. Besides
the ability to toggle the hardware prefetchers, likwid-features also reports on the
state of switchable processor features like, e.g., Intel Speedstep:
\begin{lstlisting}
$ likwid-features
-------------------------------------------------------------
CPU name:       Intel Core 2 65nm processor 
CPU core id:    0 
-------------------------------------------------------------
Fast-Strings:                   enabled
Automatic Thermal Control:      enabled
Performance monitoring:         enabled
Hardware Prefetcher:            enabled
Branch Trace Storage:           supported
PEBS:                           supported
Intel Enhanced SpeedStep:       enabled
MONITOR/MWAIT:                  supported
Adjacent Cache Line Prefetch:   enabled
Limit CPUID Maxval:             disabled
XD Bit Disable:                 enabled
DCU Prefetcher:                 enabled
Intel Dynamic Acceleration:     disabled
IP Prefetcher:                  enabled
-------------------------------------------------------------
\end{lstlisting}
Disabling, e.g., adjacent cache line prefetch then works as follows:
\begin{lstlisting}[basicstyle=\small\ttfamily]
$ likwid-features %-u CL\_PREFETCHER%
[...]
CL_PREFETCHER:  disabled
\end{lstlisting}
likwid-features currently only works for Intel Core 2 processors,
but support for other architectures is planned for the future.

\section{Comparison with PAPI}\label{sec:papi}
\label{sec:comp_papi}
\begin{table*}[t]
    \centering\renewcommand{\arraystretch}{1.2}
    \begin{tabular}{l|p{6cm}|p{6cm}}
      &LIKWID&PAPI\\
      \hline
      Dependencies & 
      Needs system headers of Linux 2.6 kernel. No other external dependencies.& 
      Needs kernel patches depending on platform and architecture. No patches necessary 
      on Linux kernels $>2.6.31$.\\
      Installation & 
      Build system based on make only. Install documentation 10 lines.  Build 
      configuration in a single text file (21 lines). & 
      Install documentation is 582 lines (3.7.2) and 397 lines (4.0.0).
      The installation of PAPI for this comparison was not without problems.\\
      Command line tools &
      Core is a collection of command line tools which are intended to be used standalone. &
      Collection of small utilities. These utilities are not supposed to be used as 
      standalone tools. There are many PAPI-based tools available from other sources.\\
      User API support &
      Simple API for configuring named code regions. API only turns counters on and off. 
      Configuration of events and output of results is still based on the command line tool. &
      Comparatively high-level API. Events must be configured in the code.\\
      Library support &
      While it can be used as library this was not initially intended. &
      Mature and well tested library API for building own tooling.\\
      Topology information &
      Listing of thread and cache topology. Results are extracted from cpuid and presented 
      in and accessible way as text and ASCII art. Nondata caches are omitted. No output of 
      TLB information. & 
      Information also based on cpuid. Utility outputs all caches (including TLBs). No 
      output of shared cache information. Thread topology only as accumulated counts of 
      HW threads and Cores. No mapping from processor Ids to thread topology.\\
      Thread and process pinning &
      There is a dedicated tool for pinning processes and threads in a portable and simple 
      manner. This tool is intended to be used together with likwid-perfCtr &
      No support for pinning.\\
      Multicore support &
      Multiple cores can be measured simultaneously. Binding of threads or processes to 
      correct cores is the responsibility of the user. &
      No explicit support for multicore measurements.\\
      Uncore support &
      Uncore events are handled by applying socket locks, which prevent multiple 
      measurements in threaded mode. &
      No explicit support for measuring shared resources.\\
      Event abstraction &
      Preconfigured event sets (so-called event groups) with derived metrics. &
      Abstraction through papi\_events, which map to native events.\\
      Platform support &
      Supports only x86-based processors on Linux with 2.6 kernel. &
      Supports a wide range of architectures on various platforms (dedicated support for 
      HPC systems like BlueGene or Cray XT3/4/5) with various operating systems 
      (Linux, FreeBSD, and Windows).\\
      Correlated measurements &
      LIKWID can measure performance counters only &
      PAPI-C can be extended to measure and correlate various data like, e.g., 
      fan speeds or temperatures.\\
    \end{tabular}
    \caption{Comparison between LIKWID and PAPI}
    \label{tab:papi_comp}
\end{table*}
PAPI \cite{PAPI-C} is a popular and well known framework to measure hardware performance
counter data. In contrast to LIKWID it relies on other software to implement
the architecture-specific parts and concentrates on providing a portable
interface to performance metrics on various platforms and
architectures. PAPI is mainly intended to be used as a library but also
includes a small collection of command line utilities. At the time of writing
PAPI is available in a classic version (PAPI 3.7.2) and a new main
branch (PAPI 4.0.0). Both version rely on autoconf to generate the build configuration. 
    
Table \ref{tab:papi_comp} compares PAPI with LIKWID without any claim for completeness.
Of course many issues are difficult to quantify, and a thorough
coverage of these points is beyond the scope of this paper. Still the
comparison should give an impression about the differences between both tools.
The most important difference is that LIKWID main focus is in providing a collection of
command line tools for the end user while PAPI's main focus is to be used as a library
by other tools.

\section{Case studies}\label{sec:cases}

\subsection{Case study 1: Influence of thread topology on STREAM triad performance}
\label{sec:case_1}

\begin{figure}[htbp]\centering
    \includegraphics*[width=0.9\linewidth]{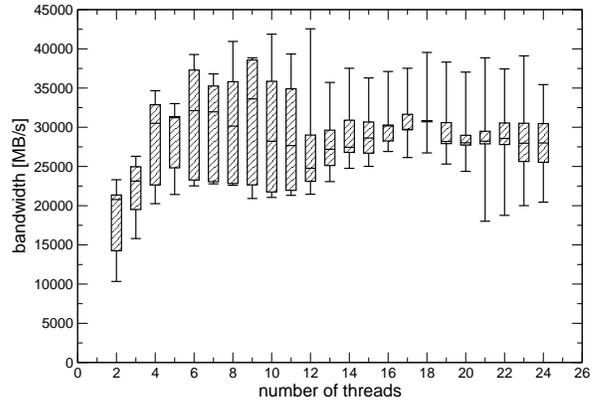}
    \caption{STREAM triad test run for the Intel icc compiler on a two-socket 
      12-core Westmere system with 100 samples per
    thread count (this will be the same for all subsequent test runs). The
    application is not explicitly pinned.  The box plot shows the 25-50 range
    with the median line.}
    \label{fig:stream_icc}
\end{figure}
\begin{figure}[htbp]\centering
    \includegraphics*[width=0.9\linewidth]{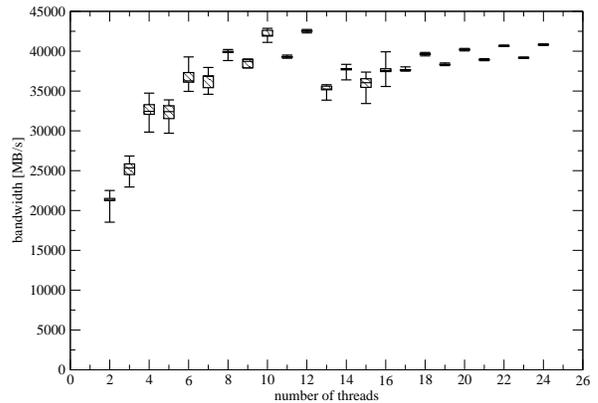}
    \caption{STREAM triad test run for the Intel icc compiler.
    The application is pinned such that threads are equally
    distributed on the sockets to utilize the memory bandwidth in the most
    effective way. Moreover the threads are first distributed over physical
    cores and then over SMT threads.}
    \label{fig:stream_icc_pinned}
\end{figure}
\begin{figure}[htbp]\centering
    \includegraphics*[width=0.9\linewidth]{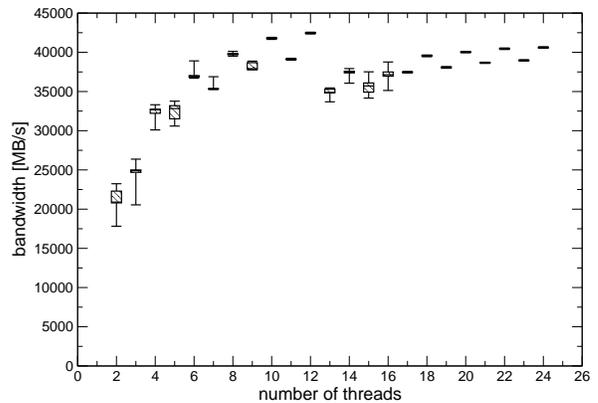}
    \caption{STREAM triad test run for the Intel icc compiler. The application
    was run with the affinity interface of the Intel OpenMP implementation set
    to ``scatter.''}
    \label{fig:stream_icc_scatter}
\end{figure}
To illustrate the general importance of thread affinity we use the well known OpenMP
STREAM triad on an Intel Westmere dual-socket system. Intel Westmere is a
hexacore design based on the Nehalem architecture and supports two SMT threads
per physical core. Two different compilers are considered: Intel icc (11.1,
with options \verb+-openmp+ \verb+-O3+ \verb+-xSSE4.2+ \verb+-fno-fnalias+) 
and gcc (4.3.3, with
options \verb+-O3+ \verb+-fopenmp+ \verb+-fargument-noalias+). 
The executable for the test on
AMD Istanbul was compiled with Intel icc (11.1, \verb+-openmp+ \verb+-O3+ 
\verb+-fno-fnalias+). 
Intel compilers support thread affinity only if the application
is executed on Intel processors. The functionality of this topology interface
is controlled by setting the environment variable \verb+KMP_AFFINITY+. In our
tests \verb+KMP_AFFINITY+ was set to \verb+disabled+.  For the case of the
STREAM triad on these ccNUMA architectures the best performance is achieved if
threads are equally distributed across the two sockets.

Figure~\ref{fig:stream_icc} shows the results for the Intel compiler with no
explicit pinning. In contrast, the data in Fig.~\ref{fig:stream_icc_pinned} 
was obtained with the threads distributed in a round-robin manner
across physical sockets using likwid-pin. As described earlier, the Intel OpenMP
implementation creates \verb+OMP_NUM_THREADS+ in addition to the initial master
thread, but the first newly created thread is used
as a ``shepherd'' and must not be pinned. likwid-pin provides a type parameter to
indicate the OpenMP implementation and automatically sets an appropriate skip
mask.  In contrast, gcc OpenMP only creates \verb+OMP_NUM_THREADS-1+ 
additional threads and does not require a shepherd thread.
As can be seen in 
Fig.~\ref{fig:stream_icc}, the non-pinned runs show a large variance in performance
especially for the smaller thread counts where the probability is large that
only one socket is used. With larger thread counts there is a high probability
that both sockets are used, still there is also a chance that cores are
oversubscribed and performance is thereby reduced. The pinned case consistently
shows high performance.

The effectiveness of the affinity functionality of the Intel OpenMP
implementation 
can be
seen in Fig.~\ref{fig:stream_icc_scatter}. This option provides
the same high performance as with likwid-pin, at all thread counts.

\begin{figure}[htbp]\centering
    \includegraphics*[width=0.9\linewidth]{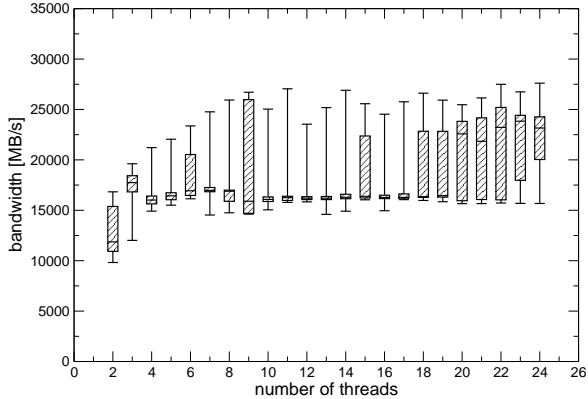}
    \caption{STREAM triad test run for the gcc compiler without pinning.}
    \label{fig:stream_gcc}
\end{figure}

\begin{figure}[htbp]\centering
    \includegraphics*[width=0.9\linewidth]{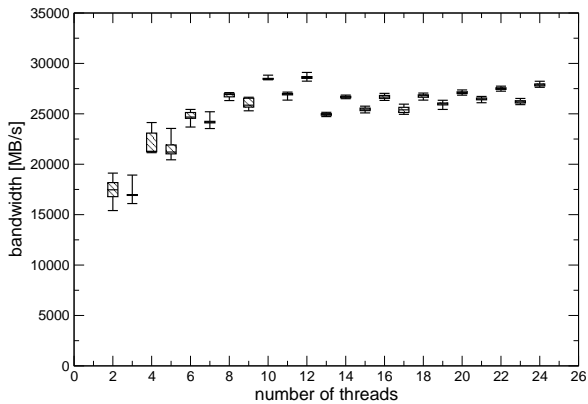}
    \caption{STREAM triad test run for the gcc compiler.
   The application was pinned with likwid-pin. The arguments for likwid-pin and the
   plot properties are the same as for the Intel icc test in 
   Fig.~\ref{fig:stream_icc_pinned}.}
    \label{fig:stream_gcc_pinned}
\end{figure}

In Fig.~\ref{fig:stream_gcc} and Fig.~\ref{fig:stream_gcc_pinned} the same
test is shown for gcc. Interestingly, the performance distribution is
significantly different compared to the non-pinned Intel icc test case in 
Fig.~\ref{fig:stream_icc}.  While with Intel icc the variance was larger for smaller
thread counts, for gcc the variance for this region is small and results are
bad with high probability. For larger thread counts this picture is reversed:
Intel icc has a small variance while gcc shows the biggest variance. One
possible explanation is that the gcc code is less dense in terms of cycles per
instruction, tolerating an oversubscription, and can probably benefit from SMT
threads to a larger extent than the Intel icc code. This behavior was not
investigated in more detail here.

\begin{figure}[htbp]\centering
    \includegraphics*[width=0.9\linewidth]{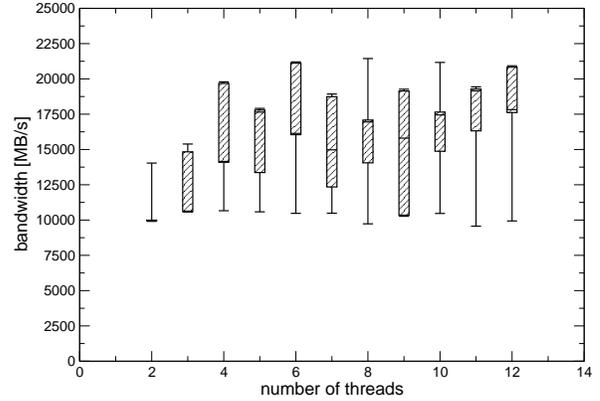}
    \caption{STREAM triad test run for the Intel icc compiler on an AMD Istanbul node
      without pinning.}
    \label{fig:stream_amd}
\end{figure}

\begin{figure}[htbp]\centering
    \includegraphics*[width=0.9\linewidth]{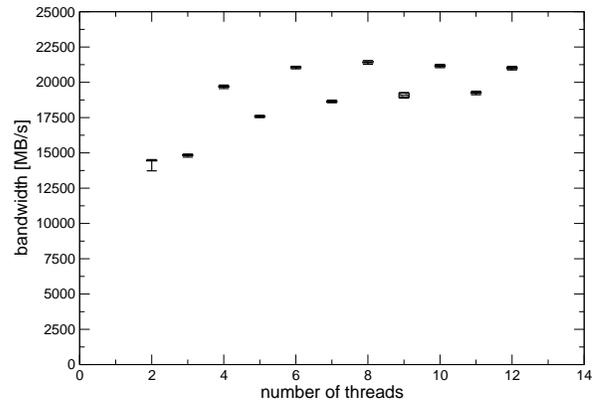}
    \caption{STREAM triad test run for the Intel icc compiler on a AMD Istanbul node.
    The application was pinned with likwid-pin. The arguments for likwid-pin are the same as in Fig.~\ref{fig:stream_icc_pinned}.}
    \label{fig:stream_amd_pinned}
\end{figure}

Finally the Intel icc executable was also benchmarked on a two-socket AMD
Istanbul hexacore node.  Fig.~\ref{fig:stream_amd} shows that there is a
large performance variance in the unpinned case, as expected. Still no
significant difference can be seen between the distribution for smaller or
larger thread counts. Enforcing affinity with  likwid-pin
(Fig.~\ref{fig:stream_amd_pinned}) yields good, stable results for all thread
counts. It is apparent that the SMT threads of Intel Westmere increase the
probability for interference of competing processes. It also makes Intel
Westmere more sensitive to oversubscription and  leads to volatile
performance with smaller thread counts.

\subsection{Case Study 2: Influence of thread topology on a topology-aware stencil code}
\label{sec:case_2}

While in the first case study the ccNUMA characteristics of the benchmark systems
only required the distribution of threads across cores to be ``uniform,'' 
the following example will show that the specific thread
and cache topology must sometimes be taken into account, and the exact
mapping of threads to cores becomes vital for getting good 
performance.

\begin{figure}[htbp]\centering
    \includegraphics*[width=0.9\linewidth]{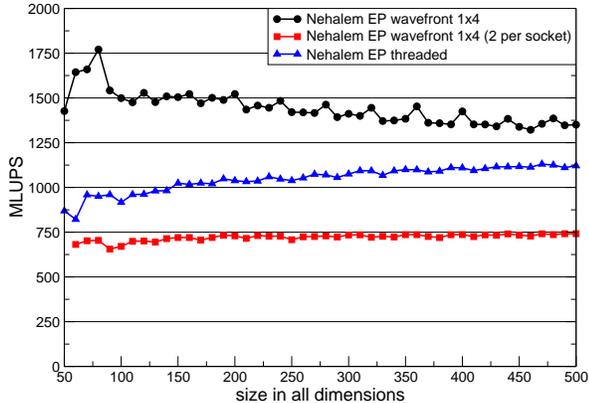}
	\caption{Performance of an optimized 3D Jacobi smoother versus
          linear problem size (cubic computational grid) on a dual-socket Intel Nehalem EP
	node (2.66\,GHz) using one thread group consisting of four threads,
        pinned to the physical cores of one socket
        (circles). In contrast, pinning pairs of threads to different sockets
        (squares) is hazardous for performance. The
	threaded baseline with nontemporal stores is shown for reference
        (triangles). Results are in million lattice site updates per second 
        [MLUPS].}
    \label{fig:wavefront_pin}
\end{figure}

We investigated a highly optimized application that
was specifically designed to utilize the shared caches of modern multicore
architectures. It implements an iterative 3D Jacobi smoother using
a 7-point stencil and is based on the POSIX threads library. All
critical computational kernels are implemented in assembly language. This code
uses implicit temporal blocking based on a pipeline parallel processing
approach~\cite{jan10}. The benchmarks were performed
on a dual-socket Intel Nehalem EP quad-core system. 
Figure~\ref{fig:wavefront_pin} shows that in case of wrong pinning the
effect of the optimization is reversed and performance is 
reduced by a factor of two, because the shared cache cannot be leveraged to increase
the computational intensity. 
In this case performance is even lower than with a naive
threaded baseline code without temporal blocking. Hence, just
pinning threads ``evenly'' through the machine is not sufficient
here; the topology of the machine requires a very specific
thread-core mapping for the blocking optimizations to become
effective.
\subsection{Case Study 3: Examining the effect of temporal blocking}
\label{sec:case_3}

Using the code from the case study in Sec.~\ref{sec:case_2}, we performed hardware
performance counter measurements with likwid-perfCtr on a dual-socket
Nehalem~EP system to quantify the effect of a temporal blocking optimization. The
measurements use three versions of a 7-point stencil Jacobi kernel: (i) a
standard threaded code with temporal stores (``threaded''), 
(ii) the same threaded
implementation with nontemporal stores (``threaded (NT)''), and
(iii) the temporal blocking code mentioned in the
previous section. The data transfer volume to and from main
memory is used as a metric to
evaluate the effect of temporal blocking. Two uncore events
are relevant here: The number of cache lines allocated in L3, and the number of
cache lines victimized from L3 (see Tab.\ref{tab:perfCtr}). In all cases, the same
number of stencil updates was executed with identical settings, and the four
physical cores of one socket were utilized. The results are shown in
Table~\ref{tab:perfCtr}. It can be seen that nontemporal stores save about
1/3 of the data transfer volume compared to the code with temporal
stores, because the write allocate on store misses is eliminated.
The optimized version again reduces the data transfer volume
significantly, as expected. However, the 4.5-fold overall decrease in
memory traffic does not translate into a proportional performance
boost. There are two reasons for this failure: (i) One data stream towards 
main memory cannot 
fully utilize the memory bandwidth on the Nehalem~EP, while the standard
threaded versions are able to saturate the bus. (ii) The 
performance difference between the
saturated main memory case and the L3 bandwidth for Jacobi is small (compared to
other, more bandwidth-starved designs),
which limits the performance benefit of temporal blocking on this processor.
See~\cite{whw10} for a performance model that describes those effects.
\begin{table}[b]
    \centering\renewcommand{\arraystretch}{1.2}
    \begin{tabular}{l|ccc}
        &threaded&threaded (NT)&blocked\\
        \hline
        \verb+UNC_L3_LINES_IN_ANY+&$5.91${}$\cdot${}$10^8$&$3.44${}$\cdot${}$10^8$&$1.30${}$\cdot${}$10^8$\\
        \verb+UNC_L3_LINES_OUT_ANY+&$5.87${}$\cdot${}$10^8$&$3.43${}$\cdot${}$10^8$&$1.29${}$\cdot${}$10^8$\\
        Total data volume [GB]&$75.39$&$43.97$&$16.57$\\
        Performance [MLUPS]&784&1032&1331
    \end{tabular}
    \caption{likwid-perfCtr measurements on one Nehalem EP socket, 
      comparing the standard threaded
      Jacobi solver with and without nontemporal stores with
      a temporally blocked variant.}
    \label{tab:perfCtr}
\end{table}

\section{Conclusion and Future Plans}\label{sec:conc}

LIKWID is a collection of command line applications supporting
performance-oriented software developers in their
effort to utilize today's multicore processors in an effective manner. LIKWID does not
try to follow the trend to provide yet another complex and sophisticated tooling
environment, which would be difficult to set up and would overwhelm the average user with
large amounts of data. Instead it tries to make the important functionality
accessible with as few obstacles as possible. The focus is put on
simplicity and low overhead. likwid-topology and likwid-pin enable the user to
account for the influence of thread and cache topology on performance and pin
their application to physical resources in all possible scenarios with one
single tool and no code changes.  Prototypically we have shown
the influence of thread topology and correct pinning on the example of the
STREAM triad benchmark. Moreover thread pinning and performance characteristics
were reviewed for an optimized topology-aware stencil code using likwid-perfCtr. 
LIKWID is open source and released under GPL2. It can be downloaded at
http://code.google.com/p/likwid/. 

LIKWID is still in alpha stage. Near-term goals are to consolidate the current
features and release a stable version, and to include support for more processor 
types. An important feature missing in
likwid-topology is to include NUMA information in the output.  likwid-pin will
be equipped with cpuset support, so that logical core IDs may be used when
binding threads. Further goals
are the combination of LIKWID with one of the available MPI profiling
frameworks to facilitate the collection of performance counter data in MPI
programs. Most of these frameworks rely on the PAPI library at the moment. 

Future plans include applying the philosophy of LIKWID to other areas like, e.g.,
profiling (also on the assembly level) and low-level benchmarking with a tool
creating a ``bandwidth map.'' This will allow a quick overview of the cache and
memory bandwidth bottlenecks in a shared-memory node, including the ccNUMA
behavior. It is also planned to port parts of LIKWID to the Windows operating system.
On popular demand, future releases will also include support for XML output.

\section*{Acknowledgment}
We are indebted to Intel Germany for providing test systems and early access
hardware for benchmarking. Many thanks to Michael Meier, who had the basic
idea for likwid-pin, implemented the prototype, and provided many useful
thoughts in discussions. This work was supported by the Competence Network for
Scientific and Technical High Performance Computing in Bavaria (KONWIHR) under
the project ``OMI4papps.''

\end{document}